\documentstyle[sprocl]{article}

\def\Journal#1#2#3#4{{#1} {\bf #2}, #3 (#4)}

\def\NPA{{\em Nucl. Phys.} A}
\def\PRL{\em Phys. Rev. Lett.}
\def\PRC{{\em Phys. Rev.} C}
\def\PRD{{\em Phys. Rev.} D}

\def\fot{\frac{1}{2}}

\def\vep{\varepsilon}
\def\al{\alpha}
\def\be{\begin{equation}}
\def\ee{\end{equation}}
\def\bea{\begin{eqnarray}}
\def\eea{\end{eqnarray}}

\begin{document}

\title{NON-ABELIAN ANOMALY AND THE LAGRANGIAN FOR RADIATIVE MUON CAPTURE}

\author{J. SMEJKAL, E. TRUHL\'{I}K}

\address{Institute of Nuclear Physics, Czech Academy of Sciences,\\
250 68 \v{R}e\v{z} n. Prague,\\
Czech Republic\\E-mail: smejkal@ujf.cas.cz, truhlik@ujf.cas.cz}

\author{F. C. KHANNA}

\address{Department of Physics, Theoretical Physics Institute, 
University of Alberta,\\
Edmonton, Alberta, Canada, T6G 2J1\\
and\\TRIUMF, 4004 Wesbrook Mall, Vancouver, B.C., Canada, V6T 2A3\\
E-mail: khanna@phys.ualberta.ca}
\maketitle\abstracts{ We discuss an anomalous Lagrangian of the
$\pi \rho\, \omega\,a_1$ system in the presence of external electroweak
fields which is suitable for constructing the amplitude 
for the radiative muon capture by proton.}

\section{Introduction}

Recent measurement of the elementary reaction
\begin{equation}
\mu^-\,+\,p\,\longrightarrow\,n\,+\,\nu_\mu\,+\,\gamma\,,  \label{mupnnug}
\end{equation}
at TRIUMF \cite{TRIUMF} provides a value of the weak induced
pseudoscalar \mbox{constant $g_P$,}

\begin{equation}
g_P(q^2\,\approx\,0.88\,m^2_\mu)\,=\,(9.8\,\pm\,0.7\,\pm\,0.3)\,g_A(0)\,,
\label{gPexp}
\end{equation}
which exceeds its value predicted by PCAC and pion-pole dominance by
a factor $\approx\,1.5$. The analysis  of the data in \cite{TRIUMF} is
based on the radiative muon capture (RMC) amplitude \cite{FAC} obtained using
low energy theorems. We have recently presented \cite{STK} the RMC 
amplitude derived from a Lagrangian of the $\pi \rho\, \omega \,a_1$ system
which reflects the SU(2)$_L \times $SU(2)$_R$ hidden local symmetry
\cite{BKY,M,KM,STG}. This amplitude coincides in the leading order
with the amplitude obtained from the low energy theorems but these amplitudes
differ in the next order in momenta $k$ (photon) and $q$ (weak current). 

Here we present our next step in analyzing the structure of the RMC amplitude.
We start from the anomalous action for the $\pi \rho\, \omega\,a_1\,D$ system
given by Kaiser and Meissner \cite{KM}.
The most general Wess-Zumino anomalous action involving pseudoscalars,
vectors, axial vectors and electroweak particles 
\newline
reads \cite{KM},
\bea
\Gamma_{an}[\xi_L,\xi_R,\xi_M,L,R,{\cal L},{\cal R}]\,& = &\,
\Gamma^{cov}_{WZW}[U,{\cal L},{\cal R}] \nonumber  \\
& & \,+\sum_{i=1}^{14}\,\int_{M^4}\,c_i\,
{\cal L}_i[\xi_L,\xi_R,\xi_M,L,R,{\cal L},{\cal R}]\,. \label{GAN}
\eea
Here $\Gamma^{cov}_{WZW}[U,{\cal L},{\cal R}]$ is the covariant 
Wess-Zumino-Witten action containing pseudoscalars and the electroweak
fields. It already satisfies the anomaly constraints. Generally, the 
14 independent
terms in the r.\,h.\,s.\, of Eq.\,(\ref{GAN}) are given in
Eqs.\,(3.8) and Eqs.\,(3.9) of Ref.\,\cite{KM}. 
As the terms ${\cal L}_1$-${\cal L}_8$ contain at least 4 particles in
each vertex, only the terms ${\cal L}_9$-${\cal L}_{14}$ are of interest
for our purpose. Later on, Kaiser and Meissner drop the weak interaction
and also the D meson. In contrast to it, we consider the full electroweak
interaction.
As a result, our anomalous Lagrangian 
contains all the natural parity violating vertices which are necessary to
construct a contribution to the RMC amplitude in such a way that 
this particular amplitude satisfies gauge invariance, CVC and PCAC 
constraints by itself.

\section{Contribution from the Wess-Zumino-Witten anomalous action}

The covariant Wess-Zumino-Witten anomalous action of pseudoscalars 
\newline reads \cite{BKY,M,W}
\be
\Gamma^{cov}_{WZW}[U,{\cal L},{\cal R}]\,=\,-i\frac{N_c}{240\,\pi^2}\,
\int_{M^5}\,{\it Tr}[\al^5]_{covariantized}\,,  \label{Gcov}
\ee
where $N_c$ is the number of colours and $\al$ is a differential
one-form
\be
\al\,=\,(\partial_\mu U) U^{\dagger} dx_\mu\,,\qquad\,
U(x)\,=\,exp[-i\Pi^a(x)\tau^a/f_\pi]\,\equiv\,\xi^2\,.
\ee
The integral is over a five-dimensional manifold $M^5$ whose boundary is 
the ordinary Minkowski space $M^4$. In the process of covariantization, one adds
terms to the non-covariantized anomalous action which contain external gauge
fields ${\cal R}_\mu$ and ${\cal L}_\mu$ in such a way that the 
covariantized form would satisfy the Wess-Zumino anomaly equation \cite{BKY}
\be
\delta\,\Gamma^{cov}_{WZW}[U,{\cal L},{\cal R}]\,=\,-\frac{N_c}{24\pi^2}\,
\int_{M^4}\,{\it Tr}[\epsilon_L\{(d{\cal L})^2\,+\,\frac{1}{2}i\,(d{\cal L})^3\}]
\,-\,(L\,\leftrightarrow\,R)\,.
\ee
The external fields ${\cal L}_\mu$ and ${\cal R}_\mu$ are directly related
to the external gauge bosons of the Standard Model \cite{BKY}.

The contribution from the action (\ref{Gcov}) to the 3-point Lagrangian 
of interest is
\be
[{\cal L}_{WZW}]^{[3]}\,=\,i\frac{e^2}{8\pi^2 f_\pi}\, \vep_{\kappa \lambda
 \mu \nu}
(\partial_\kappa \widetilde {\cal B}_\lambda)(\partial_\mu \vec {\cal V}_\nu
\,\cdot\,\vec{\Pi})\,,   \label{LWZW}
\ee
where ${\widetilde {\cal B}}_\lambda$ is an electroweak neutral field and
$\vec {\cal V}_\nu$ is a weak vector field (see below).

\section{Contribution from the homogenous terms}

As we have already mentioned, we include both the electromagnetic and 
the weak interactions and put $D\,=\,0$.
Then our analogue of Eqs.\,(3.10) of Ref.\,\cite{KM} is
\bea
\omega_L\,&=&\,\fot g\, \omega\,+\,\fot(\beta_A-\beta_V)\,-\,\frac{1}{6}\,
e\,\widetilde {\cal B}\,,  \label{OL} \\
\omega_R\,&=&\,\fot g\, \omega\,-\,\fot(\beta_A+\beta_V)\,-\,\frac{1}{6}\,
e\,\widetilde {\cal B}\,,  \label{OR} \\
\omega_M\,&=&\,\beta_M\,,  \label{OM}
\eea
\bea
F_L\,+\,\xi_M\,F_R\,\xi^\dagger_M\,&=&\,g\,d\omega\,+\,F_V\,,  \label{FLP} \\
\xi_L\,{\cal F}_L\,\xi^\dagger_L\,+\,\xi_M\,\xi_R\,{\cal F}_R\,\xi^\dagger_R\,
\xi^\dagger_M\,&=&\,\frac{1}{3}\,e\,d\widetilde {\cal B}\,+\,
\widetilde{\cal F}_{\cal V}\,,  \label{CFLP} \\
F_L\,-\,\xi_M\,F_R\,\xi^\dagger_M\,&=&\,-F_A\,,  \label{FLM} \\
\xi_L\,{\cal F}_L\,\xi^\dagger_L\,-\,\xi_M\,\xi_R\,{\cal F}_R\,\xi^\dagger_R\,
\xi^\dagger_M\,&=&\,-\widetilde{\cal F}_{\cal A}\,.  \label{CFLM}
\eea
The isoscalar and isovector components are separated as
\bea
\beta_A\,&=&\,-g\,\vec \tau \cdot \vec a\,+\,e\,\vec \tau \cdot 
\vec {\widetilde {\cal A}}\,\equiv\,-g\,a\,+\,e\,\widetilde {\cal A}\,,  \label{BA}  \\
\beta_V\,&=&\,-g\,\vec \tau \cdot \vec \rho\,+\,e\,\vec \tau \cdot 
\vec {\widetilde {\cal V}}\,\equiv\,-g\,\rho\,+\,e\,\widetilde {\cal V}\,,  \label{BV}  \\
\beta_M\,&=&\,-g\,\vec \tau \cdot \vec a\,,  \label{BM}  \\
F_V\,&=&\,g\,\vec \tau \cdot d\vec \rho\,+\,\fot i g^2\,[\,
(\vec \tau \cdot \vec \rho)(\vec \tau \cdot \vec \rho)\,+\,
(\vec \tau \cdot \vec a)(\vec \tau \cdot \vec a)\,]\,,  \\
F_A\,&=&\,g\,\vec \tau \cdot d\vec a\,+\,\fot i g^2\,[\,(\vec \tau \cdot \vec a)\,
(\vec \tau \cdot \vec \rho)\,+\,(\vec \tau \cdot \vec \rho)\,
(\vec \tau \cdot \vec a)\,]\,.  
\eea
We also introduce the following quantities,
\bea
{\widetilde {\cal F}}_{\cal V}\,&=&\,e\,
(\vec \tau \cdot d\vec {\widetilde  { \cal V}}) 
\,+\,\fot i e^2\,[\,(\vec \tau \cdot \vec {\widetilde  { \cal V}})
(\vec \tau \cdot \vec {\widetilde  { \cal V}})
 \,+\,(\vec \tau \cdot \vec {\widetilde  { \cal A}})
(\vec \tau \cdot \vec {\widetilde  { \cal A}})\,]\,,  \\
\widetilde  {\cal F}_{\cal A}\,&=&\,e\,
(\vec \tau \cdot d\vec {\widetilde  { \cal A}})
\,+\,\fot i e^2\,[\,(\vec \tau \cdot \vec {\widetilde  { \cal A}})
(\vec \tau \cdot \vec {\widetilde  { \cal V}})
 \,+\,(\vec \tau \cdot \vec {\widetilde  { \cal V}})
(\vec \tau \cdot \vec {\widetilde  { \cal A}})\,]\,,
\eea
where the fields $\vec {\widetilde  { \cal V}}$ and
$\vec {\widetilde  { \cal A}}$ are defined by the prescription
\bea
(\vec {\widetilde  { \cal V}}\,+\,\vec {\widetilde  { \cal A}}) \cdot
\vec {\tau}\,&=&\,\xi\,[\,(\vec {\cal V}\,+\,\vec {\cal A}) \cdot
\vec {\tau}\,]\,\xi^\dagger\,+\,\frac{2}{e}i\,(d\xi)\,\xi^\dagger\,, 
\label{VTPAT}  \\
(\vec {\widetilde  { \cal V}}\,-\,\vec {\widetilde  { \cal A}}) \cdot
\vec {\tau}\,&=&\,\xi^\dagger\,[\,(\vec {\cal V}\,-\,\vec {\cal A}) \cdot
\vec {\tau}\,]\,\xi\,-\,\frac{2}{e}i\,\xi^\dagger\,(d\xi)\,.
\label{VTMAT}
\eea
Due to the properties of the forms ${\widetilde {\cal F}}_{\cal V}$
and ${\widetilde {\cal F}}_{\cal A}$ the following equations hold
\bea
(\vec {\widetilde {\cal F}}_{\cal V}\,+\,
\vec {\widetilde {\cal F}}_{\cal A}) \cdot \vec {\tau}\,&=&\,
\xi\,[\,(\vec {\cal F}_{\cal V}\,+\,\vec {\cal F}_{\cal A}) \cdot 
\vec {\tau}\,]\,\xi^\dagger\,,  \label{FTVPFTA} \\
(\vec {\widetilde {\cal F}}_{\cal V}\,-\,
\vec {\widetilde {\cal F}}_{\cal A}) \cdot \vec {\tau}\,&=&\,
\xi^\dagger\,[\,(\vec {\cal F}_{\cal V}\,-\,\vec {\cal F}_{\cal A}) \cdot 
\vec {\tau}\,]\,\xi\,,  \label{FTVMFTA}
\eea
where the forms $\vec {\cal F}_{\cal V}$ and $\vec {\cal F}_{\cal A}$
correspond to the fields $\vec {\cal V}$ and $\vec {\cal A}$
\bea
{\cal F}_{\cal V}\,&=&\,e\,
(\vec \tau \cdot d\vec { \cal V}) 
\,+\,\fot i e^2\,[\,(\vec \tau \cdot \vec { \cal V})
(\vec \tau \cdot \vec { \cal V})
 \,+\,(\vec \tau \cdot \vec  { \cal A})
(\vec \tau \cdot \vec { \cal A})\,]\,,  \\
{\cal F}_{\cal A}\,&=&\,e\,
(\vec \tau \cdot d\vec  { \cal A})
\,+\,\fot i e^2\,[\,(\vec \tau \cdot \vec { \cal A})
(\vec \tau \cdot \vec { \cal V})
 \,+\,(\vec \tau \cdot \vec { \cal V})
(\vec \tau \cdot \vec {\cal A})\,]\,.
\eea
External fields $\vec {\cal V}$ and $\vec {\cal A}$ correspond to 
the gauge fields of \mbox{the Standard Model \cite{BKY}}
\bea
{\cal V}^{\pm}_\mu\,&=&\,-{\cal A}^{\pm}_\mu\,=\,\frac{1}{sin\,\Theta_w}\,
{\cal W}^{\pm}_\mu\,cos\,\Theta_c\,,  \label{VPM} \\
{\cal V}^3_\mu\,&=&\,{\cal B}_\mu\,+\,cotg\,(2\,\Theta_w)\,{\cal Z}_\mu\,=\,
\widetilde {\cal B}_\mu\,+\,\frac{1}{sin\,(2\,\Theta_w)}\,{\cal Z}_\mu\,,
\label{V3}  \\
{\cal A}^3_\mu\,&=&\,-\frac{1}{sin\,(2\,\Theta_w)}\,{\cal Z}_\mu\,. \label{A3}
\eea
Let us now use the 'unitary gauge' \cite{KM}. Employing our 
Eqs.\,(\ref{OL})-(\ref{CFLM}) we get analogues of Eqs.\,(3.12) \cite{KM}
for the Lagrangians ${\cal L}_9$-${\cal L}_{14}$
\bea
{\cal L}_9\,&=&\,-g\,d\omega\,{\it Tr}[\beta_V\,\beta_A]\,+\,
(g\,\omega\,-\,\frac{1}{3}e\,{\widetilde {\cal B}})
{\it Tr}[F_V\,\beta_A]\,, \label{L9}  \\
{\cal L}_{10}\,&=&\,-g\,d\omega\,{\it Tr}[-2\beta_V\,\beta_M]\,-\,
2\,(g\,\omega\,-\,\frac{1}{3}e\,{\widetilde{\cal B}})
{\it Tr}[F_V\,\beta_M]\,, \label{L10}  \\
{\cal L}_{12}\,&=&\,-\frac{1}{3}e\,d{\widetilde {\cal B}}\,
{\it Tr}[\beta_V\,\beta_A]\,+\,
(g\,\omega\,-\,\frac{1}{3}e\,{\widetilde {\cal B}})
{\it Tr}[{\widetilde {\cal F}}_{\cal V}\,\beta_A]\,, \label{L12}  \\
{\cal L}_{13}\,&=&\,-\frac{1}{3}e\,d{\widetilde {\cal B}}\,
{\it Tr}[-2\beta_V\,\beta_M]\,-\,
2\,(g\,\omega\,-\,\frac{1}{3}e\,{\widetilde {\cal B}})
{\it Tr}[{\widetilde {\cal F}}_{\cal V}\,\beta_M]\,, \label{L13}  \\
{\cal L}_{11}\,&=&\,{\cal L}_{14}\,=\,0\,.  \label{L11L14}
\eea
In order to be compatible with the final result of Ref.\,\cite{KM}
given in Eqs.\,(A.1) of appendix A 
we define new linear combinations 
\bea
\widetilde {\cal L}_7\,&=&\,{\cal L}_9\,+\,\fot\,{\cal L}_{10}\,, \label{LT7} \\
\widetilde {\cal L}_8\,&=&\,\fot\,{\cal L}_9\,-\,\frac{1}{4}\,{\cal L}_{10}\,, 
\label{LT8} \\
\widetilde {\cal L}_9\,&=&\,{\cal L}_{12}\,+\,\fot\,{\cal L}_{13}\,, 
\label{LT9} \\
\widetilde {\cal L}_{10}\,&=&\,\fot\,{\cal L}_{12}\,-\,\frac{1}{4}\,
{\cal L}_{13}\,. \label{LT10} 
\eea
We now
\begin{enumerate}
\item put Eqs.\,(\ref{L9})-(\ref{L13}) into Eqs.\,(\ref{LT7})-(\ref{LT10}),
\item introduce Eqs.(\ref{BA})-(\ref{BM}) into the new equations 
for $\widetilde {\cal L}_7$-$\widetilde {\cal L}_{10}$,
\item because we need only the 3-particle terms of the invariants
$\widetilde {\cal L}_7$-$\widetilde {\cal L}_{10}$, we take into account
only the linear parts of $\widetilde {\cal V}$, $\widetilde {\cal A}$ and
${\widetilde {\cal F}}_{\cal V}$ using Eqs.\,(\ref{VTPAT})-(\ref{FTVMFTA})
\be
[{\widetilde {\cal F}}_{\cal V}]^l\,=\,{\cal F}_{\cal V}\,,  
\ee
\be
[{\widetilde {\cal V}}]^l\,=\,\cal V\,,  \\
\ee
\be
[{\widetilde {\cal A}}]^l\,=\,{\cal A}\,+\,\frac{i}{e}\,[(d\xi)\,
\xi^\dagger\,+\,\xi^\dagger\,(d\xi)]^l 
\,=\,{\cal A}\,+\,\frac{1}{ef_\pi}\,d\Pi\,, 
\ee
Here $Q\,\equiv\,\vec {Q} \cdot \vec{\tau}$ for $Q\,=\,{\cal V},\,{\cal A},\,
\Pi$.
\item remove the non-physical $\pi-a_1$ mixing by the redefinition of the
axial meson field \cite{BKY,STG}
\be
\tilde a\,=\,a-\frac{1}{2f_\pi\,g}\,d\Pi\,.
\ee
\end{enumerate}
As a result, we have the following set of equations
\bea
[{\widetilde {\cal L}}_7]^{[3]}\,&=&\,g\,d\omega\,{\it Tr}[(g\rho-e{\cal V})
(\frac{1}{f_\pi}\,d\Pi\,+\,e{\cal A})] \nonumber  \\
& & +\,(g\omega\,-\,\frac{1}{3}e\,\widetilde {{\cal B}})\,
{\it Tr}[F_V(\frac{1}{f_\pi}\,d\Pi\,+\,e{\cal A})]\,,  \label{LT73}
\eea
\bea
[\widetilde {{\cal L}}_8]^{[3]}\,&=&\,-g\,d\omega\,{\it Tr}[(g\rho-e{\cal V})
(g{\tilde a}\,-\,\fot e{\cal A})] \nonumber  \\
& & -\,(g\omega\,-\,\frac{1}{3}e\,{\widetilde {\cal B}})\,
{\it Tr}[F_V(g{\tilde a}\,-\,\fot e{\cal A})]\,,  \label{LT83}
\eea
\bea
[{\widetilde {\cal L}}_9]^{[3]}\,&=&\,+\frac{1}{3}e\,d{\widetilde {\cal B}}\,
{\it Tr}[(g\rho-e{\cal V})
(\frac{1}{f_\pi}\,d\Pi\,+\,e{\cal A})] \nonumber  \\
& & +\,(g\omega\,-\,\frac{1}{3}e\,\widetilde {{\cal B}})\,
{\it Tr}[{\cal F}_{\cal V}(\frac{1}{f_\pi}\,d\Pi\,+\,e{\cal A})]\,,  \label{LT93}
\eea
\bea
[\widetilde {{\cal L}}_{10}]^{[3]}\,&=&\,-\frac{1}{3}e\,d{\widetilde {\cal B}}\,
{\it Tr}[(g\rho-e{\cal V})
(g{\tilde a}\,-\,\fot e{\cal A})] \nonumber  \\
& & -\,(g\omega\,-\,\frac{1}{3}e\,{\widetilde {\cal B}})\,
{\it Tr}[{\cal F}_{\cal V}(g{\tilde a}\,-\,\fot e{\cal A})]\,.  \label{LT103}
\eea
We can now compare our Eqs.\,(\ref{LT73})-(\ref{LT103}) with Eqs.\,(A.1) of
Ref.\,\cite{KM}. It is easy to see that our $\widetilde {{\cal L}}_7$
and $\widetilde {{\cal L}}_8$ coincide with the same Lagrangians derived
in \cite{KM}, if we put the external fields to zero. In this approximation, 
the Lagrangians $\widetilde {{\cal L}}_9$ and $\widetilde {{\cal L}}_{10}$
are also zero. Naturally they are absent in Ref.\,\cite{KM}.

The parametrization of the interactions in the $\pi \rho\,\omega\, a_1$ system
in the form $\widetilde {{\cal L}}_7$-$\widetilde {{\cal L}}_{10}$ has 
the advantage that each of these Lagrangians leads to the currents satisfying
PCAC by itself. But in this case, the field of the $a_1$ meson enters
isolated from the pion field, while the external gauge field $\cal A$
enters simultaneously with the fields of both mesons. It will be more 
suitable for our purpose to separate the $a_1$ and $\cal A$ fields.
This is accomplished in the new combinations of the Lagrangians
$\widetilde {{\cal L}}_7$-$\widetilde {{\cal L}}_{10}$
\be
{\bar {\cal L}}_i\,=\,{\widetilde {\cal L}}_i\,,\qquad\,i=7,9\,,  
\ee
\be
{\bar {\cal L}}_j\,=\,{\widetilde {\cal L}}_j\,
-\,\fot{\widetilde {\cal L}}_{j-1}\,,\qquad\,j=8,10\,.
\ee
We now present the final result written in the form convenient for
construction of the amplitude for the radiative muon capture. With
the help \mbox{ of Eqs.\,(A.2) \cite{KM} } we get
\bea
[\bar {\cal L}_7]^{[3]}\,&=&\,2ig\,\vep_{\kappa \lambda \mu \nu}\,\{
\partial_\kappa \omega_\lambda\,[( g \vec {\rho}_\mu\,-\,e\vec {\cal V}_\mu)
\cdot
(\frac{1}{f_\pi}\,\partial_\nu\vec {\Pi}\,+\,e\vec {\cal A}_\nu)] 
\nonumber  \\
& & +\,(g\omega_\kappa\,-\,\frac{1}{3}e\,\widetilde {\cal B}_\kappa)\,
[(\partial_\lambda \vec {\rho}_\mu) \cdot (\frac{1}{f_\pi}\,
\partial_\nu \vec {\Pi}\,+\,e\vec {\cal A}_\nu)]\}\,,  \label{LB73}
\eea
\bea
[\bar {\cal L}_8]^{[3]}\,&=&\,-2ig\,\vep_{\kappa \lambda \mu \nu}\,\{
\partial_\kappa \omega_\lambda\,[( g \vec {\rho}_\mu\,-\,e\vec {\cal V}_\mu)
\cdot
(g \vec {\tilde a}_\nu\,+\,\frac{1}{2f_\pi} \partial_\nu \vec {\Pi})] 
\nonumber  \\
& & +\,(g\omega_\kappa\,-\,\frac{1}{3}e\,{\widetilde {\cal B}}_\kappa)\,
[(\partial_\lambda \vec {\rho}_\mu) \cdot (g \vec {\tilde a}_\nu\,+\,
\frac{1}{2f_\pi}\,\partial_\nu \vec {\Pi})]\}\,,  \label{LB83}
\eea
\bea
[\bar {\cal L}_9]^{[3]}\,&=&\,2ie\,\vep_{\kappa \lambda \mu \nu}\,\{
\frac{1}{3}\,\partial_\kappa {\widetilde {\cal B}}_\lambda\,
[(g \vec {\rho}_\mu\,-\,e\vec {\cal V}_\mu) \cdot
(\frac{1}{f_\pi}\,\partial_\nu \vec {\Pi}\,+\,e\vec {\cal A}_\nu)] \nonumber  \\
& & +\,(g\omega_\kappa\,-\,\frac{1}{3}e\,\widetilde {\cal B}_\kappa)\,
[(\partial_\lambda \vec {\cal V}_\mu)(\frac{1}{f_\pi}\,\partial_\nu \vec {\Pi}
\,+\,e\vec {\cal A}_\nu)]\}\,,  \label{LB93}
\eea
\bea
[\bar {{\cal L}}_{10}]^{[3]}\,&=&\,-2ie\,\vep_{\kappa \lambda \mu \nu}\,\{
\frac{1}{3}\,\partial_\kappa {\widetilde {\cal B}}_\lambda\,
[(g \vec {\rho}_\mu\,-\,e \vec {\cal V}_\mu) \cdot
(g \vec {\tilde a}_\nu\,+\,\frac{1}{2f_\pi} \partial_\nu \vec {\Pi})] 
\nonumber  \\
& & +\,(g\omega_\kappa\,-\,\frac{1}{3}e\,{\widetilde {\cal B}}_\kappa)\,
[(\partial_\lambda \vec {\cal V}_\mu)(g \vec {\tilde a}_\nu\,
+\,\frac{1}{2f_\pi} \partial_\nu \vec {\Pi})]\}\,.  \label{LB103}
\eea
The strong vertices $\rho \,\omega\,\pi$ and $\rho\,\omega\,a_1$ are already known
from the appendix A of Ref.\,\cite{KM}. 
Then Eqs.(\ref{LB73})-(\ref{LB103}) provide all other terms due to the presence of
the external electroweak interactions which violate the natural parity
and which should be used when constructing the amplitude for the radiative
muon capture by proton. Their presence guarantees that such an amplitude will
satisfy gauge invariance and CVC and PCAC constraints at the tree level exactly.
The construction of the amplitude is in progress.

\section*{Acknowledgments}
This work was supported in part by the grant GA \v{C}R 202/97/0447.
The research of F.C.K. is supported in part by NSERCC.

\end{document}